\documentclass[12pt]{amsart}
\usepackage{amssymb,latexsym}
\usepackage{epsfig}
\newtheorem{theorem}{Theorem}

\newcommand{\Rbb}{\mathbb{R}}
\newcommand{\zbar}{\overline{z}}

\newcommand{\sech}{\,\text{sech}\,}

\newcommand{\calt}{\hat{\mathcal{T}}}

\newcommand{\tilt}{\hat{\mathfrak{T}}}
\newcommand{\cre}{\hat{a}^{\dag}}
\newcommand{\anh}{\hat{a}}
\newcommand{\crx}{\hat{a}^{\dag}_{x}}
\newcommand{\anx}{\hat{a}_{x}}
\newcommand{\cry}{\hat{a}^{\dag}_{y}}
\newcommand{\any}{\hat{a}_{y}}
\newcommand{\Op}{\text{Op}^{W}_{h}}
\newcommand{\Mtw}{\tilde{M}_{\sqrt{2}}}

\begin{document}

\title[Manipulation of Semiclassical Laguerre-Gaussian Modes: a Model Case]{Manipulation of Semiclassical\\ Laguerre-Gaussian Modes: a Model Case}
\author{Michael VanValkenburgh}
\address{UCLA Department of Mathematics, Los Angeles, CA 90095-1555, USA}
\email{mvanvalk@ucla.edu}
\date{February 18, 2009}

\begin{abstract}
We continue the study, from a semiclassical viewpoint, of Calvo
and Pic\'{o}n's operators, as manipulating photon states in
quantum communication. In a previous paper, we defined a
one-dimensional model operator and studied it analytically before
moving on to Calvo and Pic\'{o}n's full two-dimensional operators.
In the present paper, we show how the one-dimensional operator may
also be useful as an experimental model, since it allows
manipulations of two-dimensional Laguerre-Gaussian modes; the
intensity distributions (in physical space) of the
Laguerre-Gaussian modes then approximately flow along the elliptic
curves studied earlier. Since the Wigner transform is fundamental
in the study of Laguerre-Gaussian modes, we give a slightly
expanded and improved treatment of the semiclassical Wigner
transform, which was only briefly mentioned in the previous paper.
\end{abstract}

\maketitle

\section{Introduction}\label{S:Intro}

Gabriel F. Calvo and Antonio Pic\'{o}n \cite{R:CalvoPicon} introduced a class of operators which allows arbitrary manipulations of the three lowest-order Hermite-Gaussian (HG) modes for the purposes of quantum communication. In our previous paper \cite{R:VVCP}, we (1) classified the self-adjoint extensions of the generators and (2) showed how to construct semiclassical approximations of the associated unitary operators; for the latter, we first considered a simple model operator to illustrate the semiclassical methods in a simplified setting before then moving on to the full operators of Calvo and Pic\'{o}n.

In this paper, we show that the simplified operator also serves as a model operator from an experimental viewpoint, when considered as manipulating Laguerre-Gaussian (LG) modes of orders $(0,0)$, $(1,0)$, $(0,1)$, and $(1,1)$.

Calvo and Pic\'{o}n introduced the following eight generators acting on Hermite-Gaussian modes:
$$\calt_{1}=\frac{1}{2}(\crx\any+\cry\anx),\qquad\calt_{2}=-\frac{i}{2}(\crx\any-\cry\anx),\qquad\calt_{3}=\frac{1}{2}(\crx\anx-\cry\any),$$
$$\calt_{4}=\frac{1}{2}(\crx+\anx-\crx\crx\anx-\crx\anx\anx-\crx\cry\any-\anx\cry\any),$$
$$\calt_{5}=-\frac{i}{2}(\crx-\anx-\crx\crx\anx+\crx\anx\anx-\crx\cry\any+\anx\cry\any),$$
$$\calt_{6}=\frac{1}{2}(\cry+\any-\cry\cry\any-\cry\any\any-\cry\crx\anx-\any\crx\anx),$$
$$\calt_{7}=-\frac{i}{2}(\cry-\any-\cry\cry\any+\cry\any\any-\cry\crx\anx+\any\crx\anx),$$
$$\calt_{8}=\frac{1}{2\sqrt{3}}[-2+3(\crx\anx+\cry\any)],$$
defined in terms of the creation and annihilation operators $\crx=\frac{1}{\sqrt{2}}\left(x-\frac{\partial}{\partial x}\right)$ and $\anx=\frac{1}{\sqrt{2}}\left(x+\frac{\partial}{\partial x}\right)$, respectively, and similarly for the $y$-variable.

These generators, within the subspace generated by the lowest three Hermite-Gaussian modes
$\mathcal{H}_{\calt}=\{|0,0\rangle,|1,0\rangle,|0,1\rangle\}$, obey the $SU(3)$ algebra
$$[\calt_{a},\calt_{b}]=if_{abc}\calt_{c}$$
($a,b,c=1,2,\ldots 8$), where the only nonvanishing (up to permutations) structure constants $f_{abc}$ are given by
$$f_{123}=1, \quad f_{147}=f_{165}=f_{246}=f_{257}=f_{345}=f_{376}=1/2,\quad\text{and }f_{458}=f_{678}=\sqrt{3}/2.$$
We note that the triad of generators
$$\Gamma_{1}\equiv\{\calt_{1},\calt_{2},\calt_{3}\}$$
gives a $SU(2)$ group that conserves the mode order. The remaining two $SU(2)$ groups are formed by the triads $$\Gamma_{2}\equiv\{\calt_{4},\calt_{5},(\calt_{3}+\sqrt{3}\calt_{8})/2\} \qquad \text{and}\qquad\Gamma_{3}\equiv\{\calt_{6},\calt_{7},(-\calt_{3}+\sqrt{3}\calt_{8})/2\}.$$
Unitary operators $\hat{U}_{\Gamma_{1}}$ generated by the first triad give rise to superpositions between the two modes $|1,0\rangle$ and $|0,1\rangle$, leaving invariant the fundamental mode $|0,0\rangle$. Unitarities $\hat{U}_{\Gamma_{2}}$ and $\hat{U}_{\Gamma_{3}}$, generated by the second and third triads, produce superpositions between the two modes $|0,0\rangle$ and $|1,0\rangle$ (leaving invariant $|0,1\rangle$), or the modes $|0,0\rangle$ and $|0,1\rangle$ (leaving invariant $|1,0\rangle$), respectively.

Here we instead consider the one-dimensional operator
\begin{equation*}
    \calt=\cre+\anh-\cre\cre\anh-\cre\anh\anh,
\end{equation*}
acting on the one-dimensional Hermite functions $h_{n}$. This is a simplified version of the operator $\calt_{4}$ above. We introduced this operator in \cite{R:VVCP} since it exhibits the same essential behavior as $\calt_{4}$, while leaving out the unessential additional variables. Here we will show that this operator may also be useful in experimental work, as an easier preliminary step before moving to the more complicated operator $\calt_{4}$.

One may check that $\calt h_{0}=h_{1}$ and $\calt h_{1}=h_{0}$. More generally,
\begin{equation*}
    \calt h_{n}=\beta_{n}h_{n+1}+\beta_{n-1}h_{n-1}
\end{equation*}
where
\begin{equation*}
    \beta_{n}=(1-n)\sqrt{n+1}.
\end{equation*}
So $\calt$, when acting on the domain of (finite) linear combinations of HG modes, behaves precisely as the infinite Jacobi matrix
\begin{equation*}
    \left(
    \begin{matrix}
    0&\beta_{0}&0&0&\cdots\\
    \beta_{0}&0&\beta_{1}&0\\
    0&\beta_{1}&0&\beta_{2}\\
    0&0&\beta_{2}&0&\ddots\\
    \vdots&&&\ddots&\ddots
    \end{matrix}
    \right)
    =
    \left(
    \begin{matrix}
    0&1&0&0&0&0&\cdots\\
    1&0&0&0&0&0\\
    0&0&0&-\sqrt{3}&0&0&\cdots\\
    0&0&-\sqrt{3}&0&-4&0\\
    0&0&0&-4&0&\ddots\\
    0&0&0&0&\ddots&\ddots&\ddots\\
    \vdots&&\vdots&&&\ddots&\ddots&\ddots
    \end{matrix}
    \right).
\end{equation*}
Using a theorem from Berezanskii's book (\cite{R:Berezanskii}, p.507; see also \cite{R:VVCP}), one can show that the deficiency index of this operator is $(1,1)$, and hence the space of boundary values is of dimension $2=1+1$. Moreover, using results of Allahverdiev \cite{R:Alla}, one can explicitly classify all self-adjoint extensions in terms of certain boundary values at infinity. This was achieved for the full operators of Calvo and Pic\'{o}n in \cite{R:VVCP}, using a slight modification of Allahverdiev's methods, but for the simplified operator $\calt$ the methods of Allahverdiev may be applied without any modification.

We will show that this model operator $\calt$ allows simple manipulations of LG modes. But first, since the Wigner transform plays a basic role in the theory of LG modes, we recall some relevant facts in the next section.

\vspace{12pt}

\section{The Semiclassical Wigner Transform}\label{S:scWig}

The standard $n$-dimensional semiclassical Wigner transform is defined as
\begin{equation}\label{E:WigStd}
    W(f,g)(x,\xi)=(2\pi h)^{-n/2}\int e^{-ip\xi/h}f(x+\frac{1}{2}p)\overline{g(x-\frac{1}{2}p)}\, dp,
\end{equation}
where $f$ and $g$ are square-integrable functions of $n$ variables. Then, among other useful properties (one may consult Folland's book \cite{R:Folland}), we have the norm-preserving property:
\begin{equation*}
    ||W(f,g)||_{L^{2}(\Rbb^{2n})}=||f||_{L^{2}(\Rbb^{n})}||g||_{L^{2}(\Rbb^{n})}.
\end{equation*}
Also, the Wigner transform has an important relationship with Weyl quantization. We recall that the semiclassical Weyl quantization of a symbol $\sigma$ is the operator $\Op(\sigma)$ defined by
\begin{equation*}
    \Op(\sigma)f(x)=(2\pi h)^{-n}\iint e^{i(x-y)\xi/h}\sigma((x+y)/2, \xi)f(y)dy\,d\xi.
\end{equation*}
Then
\begin{equation*}
    \langle \Op(\sigma)f|g\rangle =(2\pi h)^{-n/2}\iint \sigma(x,\xi)W(f,g)(x,\xi)dx\,d\xi.
\end{equation*}
As discussed in the previous paper \cite{R:VVCP}, we have an approximate formula for the evolution of the Wigner transform. Let $\tilt$ be a self-adjoint semiclassical pseudodifferential operator with Hamilton flow $\kappa_{t}$ generated by the (possibly $h$-dependent) Weyl symbol, and let $U_{t}=e^{-it\tilt/h}$ be the unitary propagator. (Note: In the previous paper, $U_{t}$ sometimes denoted the semiclassical \emph{approximate} unitary propagator.) We then have
\begin{equation*}
    W(f,g)(\kappa_{-t}(x,\xi))=W(U_{t}f,U_{t}g)(x,\xi)+\mathcal{O}(h^{2})||f||_{L^{2}(\Rbb^{n})}||g||_{L^{2}(\Rbb^{n})}.
\end{equation*}
This essentially follows from Egorov's theorem with an $\mathcal{O}(h^{2})$ error (see \cite{R:VVCP} and references therein).

In the previous paper, we briefly discussed two-dimensional Wigner transforms. In the present paper, in the next section, we will consider the case $n=1$, since LG modes are themselves one-dimensional Wigner transforms of Hermite functions \cite{R:VVLG}.

\vspace{12pt}

\section{The Semiclassical Laguerre-Gaussian Modes}\label{S:scLG}

In the context of LG modes, it is more convenient to use an alternative definition of the Wigner transform.  We define the (one-dimensional) \emph{extended} Wigner transform, of a function $F$ of two variables, as
\begin{equation*}
    \tilde{W}(F)(x,\xi)=(2\pi h)^{-1/2}\int e^{ip\xi/h}
    F\left(\frac{x+p}{\sqrt{2}},\frac{x-p}{\sqrt{2}}\right)\,
    dp.
\end{equation*}
(The sign of the phase is intentionally different from (\ref{E:WigStd}).) This is clearly a unitary operator; moreover, as shown in \cite{R:VVLG}, the $(m,n)^{th}$ LG mode is precisely given by $\tilde{W}h_{mn}$, where $h_{mn}$ is the $(m,n)^{th}$ HG mode. For reference, we will now state the main facts in the semiclassical setting.

The zeroth order HG mode is the Gaussian function
\begin{equation*}
    h_{00}(x,y)=(\pi h)^{-1/2}e^{-(x^{2}+y^{2})/(2h)}.
\end{equation*}
All other HG modes, $h_{mn}$, may be recovered by applying the creation operators
\begin{equation*}
    \hat{a}_{1}^{\dag}=(2h)^{-1/2}\left(x-h\frac{\partial}{\partial x}\right)\quad\text{and}\quad \hat{a}_{2}^{\dag}=(2h)^{-1/2}\left(y-h\frac{\partial}{\partial y}\right).
\end{equation*}
That is, $$h_{mn}=(m!n!)^{-1/2}\hat{a}_{1}^{\dag\,m}\hat{a}_{2}^{\dag\,n}h_{00}.$$
The corresponding annihilation operators are
\begin{equation*}
    \hat{a}_{1}=(2h)^{-1/2}\left(x+h\frac{\partial}{\partial x}\right)\quad\text{and}\quad \hat{a}_{2}=(2h)^{-1/2}\left(y+h\frac{\partial}{\partial y}\right).
\end{equation*}

Alternatively, to recover the LG modes, one applies to $h_{00}$ the creation operators
\begin{equation*}
    \hat{A}_{+}^{\dag}=\frac{1}{\sqrt{2}}(a_{1}^{\dag}+ia_{2}^{\dag})\quad\text{and}\quad \hat{A}_{-}^{\dag}=\frac{1}{\sqrt{2}}(a_{1}^{\dag}-ia_{2}^{\dag}).
\end{equation*}
The corresponding LG mode annihilation operators are
\begin{equation*}
    \hat{A}_{+}=\frac{1}{\sqrt{2}}(a_{1}-ia_{2})\quad\text{and}\quad \hat{A}_{-}=\frac{1}{\sqrt{2}}(a_{1}+ia_{2}).
\end{equation*}

As discussed in \cite{R:VVLG} (but also as one may easily check), the extended Wigner transform intertwines the two classes of creation and annihilation operators:
\begin{equation}\label{E:intertwine}
    \begin{aligned}
        \hat{A}_{+}^{\dag}\tilde{W} &= \tilde{W}\hat{a}_{1}^{\dag},\qquad\qquad
        \hat{A}_{-}^{\dag}\tilde{W} = \tilde{W}\hat{a}_{2}^{\dag},\\
        \hat{A}_{+}\tilde{W} &= \tilde{W}\hat{a}_{1},\qquad\qquad
        \hat{A}_{-}\tilde{W} = \tilde{W}\hat{a}_{2}.
    \end{aligned}
\end{equation}

Since $\tilde{W}h_{00}=h_{00}$, this shows that the LG modes are indeed the extended Wigner transforms of the HG modes. Moreover, one may deduce the following familiar analytic expressions for the LG modes. Suppose $x,y\in\Rbb$, and let $z=x+iy$. Then, with $h_{jk}$ denoting the $(j,k)^{th}$ HG mode, the $(j,k)^{th}$ LG mode is
\begin{equation*}
    \tilde{W}(h_{jk})(x,y)=
    \begin{cases}
        (\pi h)^{-1/2}(k!/j!)^{1/2}
        (-1)^{k}(z/\sqrt{h})^{j-k}e^{-z\zbar/(2h)}L^{j-k}_{k}(z\zbar/h)
        &\text{if }j\geq k, \text{ and}\\
        (\pi h)^{-1/2}(j!/k!)^{1/2}
        (-1)^{j}(\overline{z}/\sqrt{h})^{k-j}e^{-z\zbar/(2h)}L^{k-j}_{j}(z\zbar/h)
        &\text{if }j\leq k,
    \end{cases}
\end{equation*}
where $$L^{\alpha}_{n}(x)=\frac{x^{-\alpha}e^{x}}{n!}\frac{d^{n}}{dx^{n}}(e^{-x}x^{n+\alpha})$$
are the Laguerre polynomials. Of particular importance in this paper are $L_{0}^{0}(x)=L_{0}^{1}(x)=1$ and $L_{1}^{0}(x)=1-x$.

\vspace{12pt}

\section{Manipulation of Semiclassical LG Modes}

We first consider the action of the tensor product $\calt\otimes\calt$ on tensor products of functions:
$$(\calt\otimes\calt)(f\otimes\overline{g})=\calt(f)\otimes\overline{\calt(g)}.$$ Then we simply pull back this action to operate on extended Wigner transforms of functions. We write this as follows:
\begin{equation*}
    (\calt^{\star}\tilde{W})(f\otimes\overline{g}):=\tilde{W}(\calt(f)\otimes\overline{\calt(g)}).
\end{equation*}
Since the extended Wigner transform has the intertwining property (\ref{E:intertwine}), we see that the pullback operator $\calt^{\star}$ is precisely the differential operator
\begin{equation*}
    (\hat{A}_{+}^{\dag}+\hat{A}_{+}-\hat{A}_{+}^{\dag}\hat{A}_{+}^{\dag}\hat{A}_{+}-\hat{A}_{+}^{\dag}\hat{A}_{+}\hat{A}_{+})
    \circ(\hat{A}_{-}^{\dag}+\hat{A}_{-}-\hat{A}_{-}^{\dag}\hat{A}_{-}^{\dag}\hat{A}_{-}-\hat{A}_{-}^{\dag}\hat{A}_{-}\hat{A}_{-}).
\end{equation*}
However, this point of view only seems to complicate matters.

We are especially interested in the case when $f\otimes\overline{g}=h_{mn}$, the $(m,n)^{th}$ HG mode; then the pullback $\calt^{\star}$ acts on LG modes. In particular, $\calt\otimes\calt$ acts on the four ``binary'' HG modes as follows:
\begin{align*}
    (\calt\otimes\calt)h_{00}=h_{11},\qquad &(\calt\otimes\calt)h_{11}=h_{00},\\
    (\calt\otimes\calt)h_{10}=h_{01},\qquad &(\calt\otimes\calt)h_{01}=h_{10}.
\end{align*}
Hence
\begin{align*}
    (\calt^{\star}\tilde{W})h_{00}=\tilde{W}h_{11},\qquad &(\calt^{\star}\tilde{W})h_{11}=\tilde{W}h_{00},\\
    (\calt^{\star}\tilde{W})h_{10}=\tilde{W}h_{01},\qquad &(\calt^{\star}\tilde{W})h_{01}=\tilde{W}h_{10}.
\end{align*}

These lowest-order LG modes may be written in polar coordinates as
\begin{align*}
    \tilde{W}h_{00}&=(\pi h)^{-1/2}e^{-r^{2}/(2h)},\qquad &&\tilde{W}h_{11}=-(\pi h)^{-1/2}(1-(r/\sqrt{h})^{2})e^{-r^{2}/(2h)},\\
    \tilde{W}h_{10}&=(\pi h)^{-1/2}(r/\sqrt{h})e^{i\theta}e^{-r^{2}/(2h)},\qquad &&\tilde{W}h_{01}=(\pi h)^{-1/2}(r/\sqrt{h})e^{-i\theta}e^{-r^{2}/(2h)}.
\end{align*}
So we see that the pullback $\calt^{\star}$ interchanges the LG modes $\tilde{W}h_{00}$ and $\tilde{W}h_{11}$ and simply causes a phase reversal in the LG modes of order one.

For convenience, in order to have a semiclassical differential operator, we define
\begin{equation*}
    \tilt=-2^{-1/2}h^{3/2}\calt,
\end{equation*}
and we note that\footnote{In \cite{R:VVCP} we mistakenly studied the operator $\frac{1}{2}x(x^{2}+(hD)^{2})-\frac{3}{2}hx-\frac{1}{2}ih^{2}D$ with Weyl symbol $\frac{1}{2}x(x^{2}+\xi^{2})-\frac{3}{2}hx$, which is not quite the same as $\tilt$. However, it was merely given as an example, to illustrate a general method, so in that paper one may simply redefine the operator in the first place.}
\begin{equation*}
    \tilt=\frac{1}{2}x(x^{2}+(hD)^{2})-2hx-\frac{1}{2}ih^{2}D,
\end{equation*}
with the notation $D=\frac{1}{i}\frac{\partial}{\partial x}$.

\begin{figure}
\begin{center}

\begin{tabular}{ l c r }
  \epsfig{file=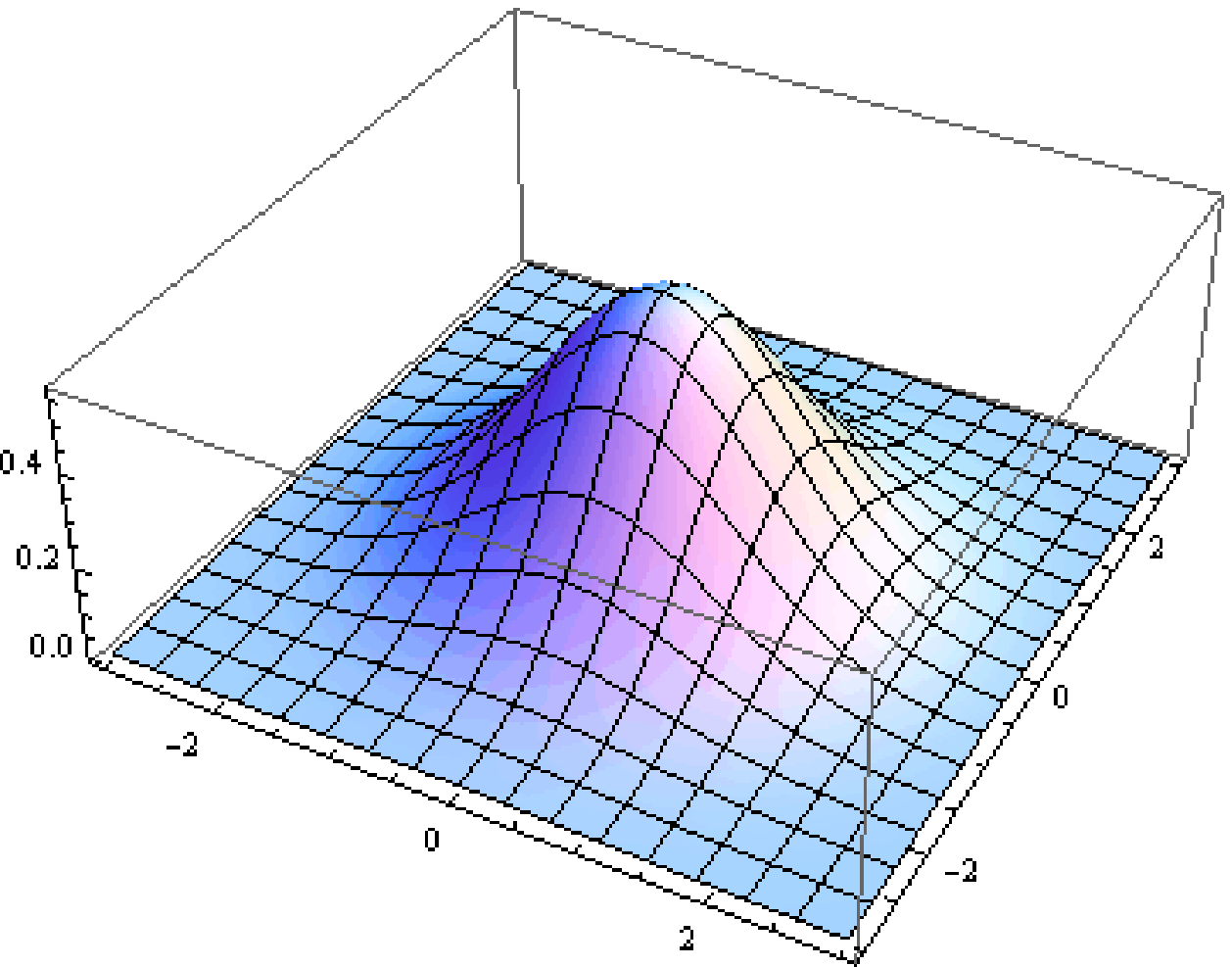,height=4cm} & \epsfig{file=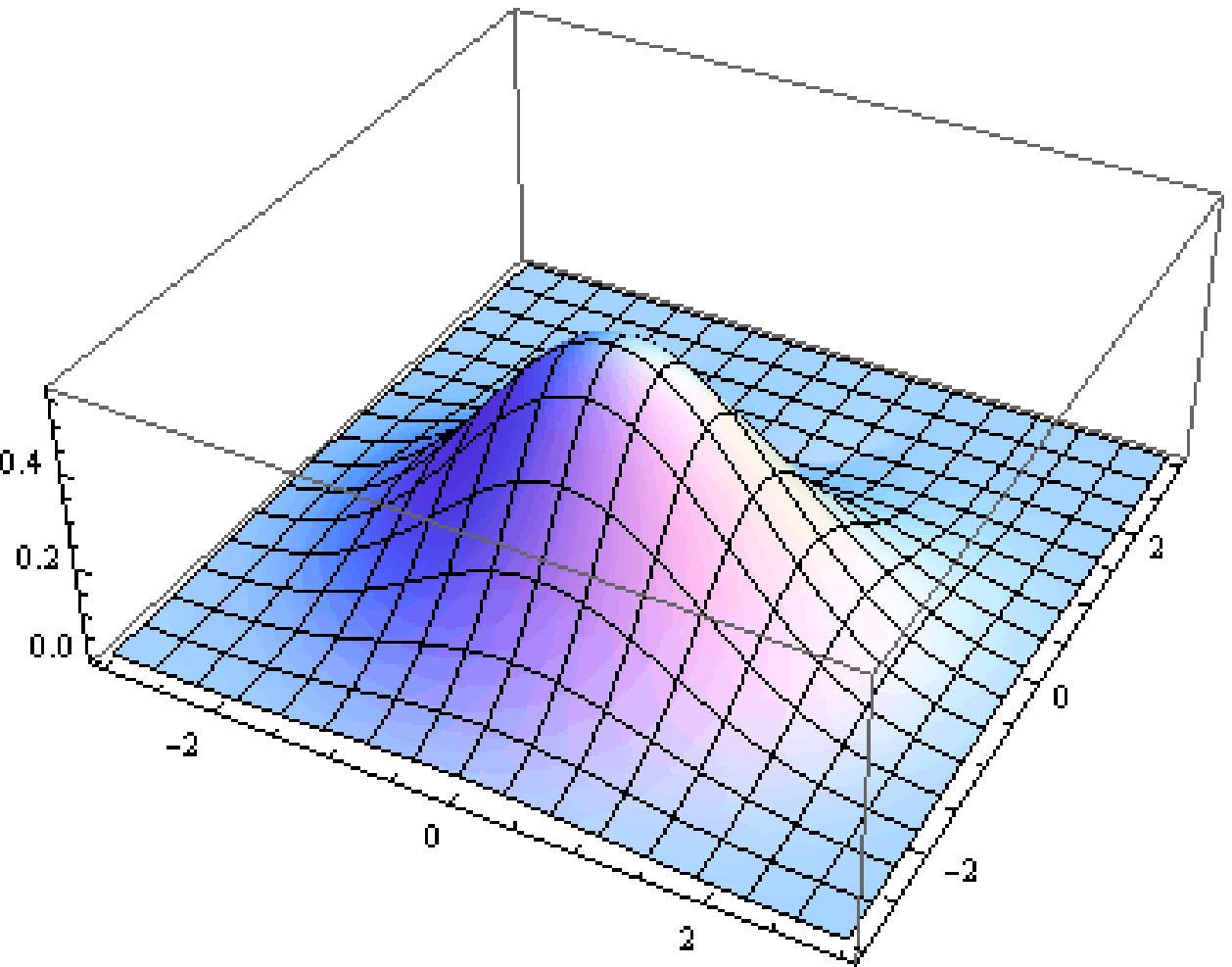,height=4cm} & \epsfig{file=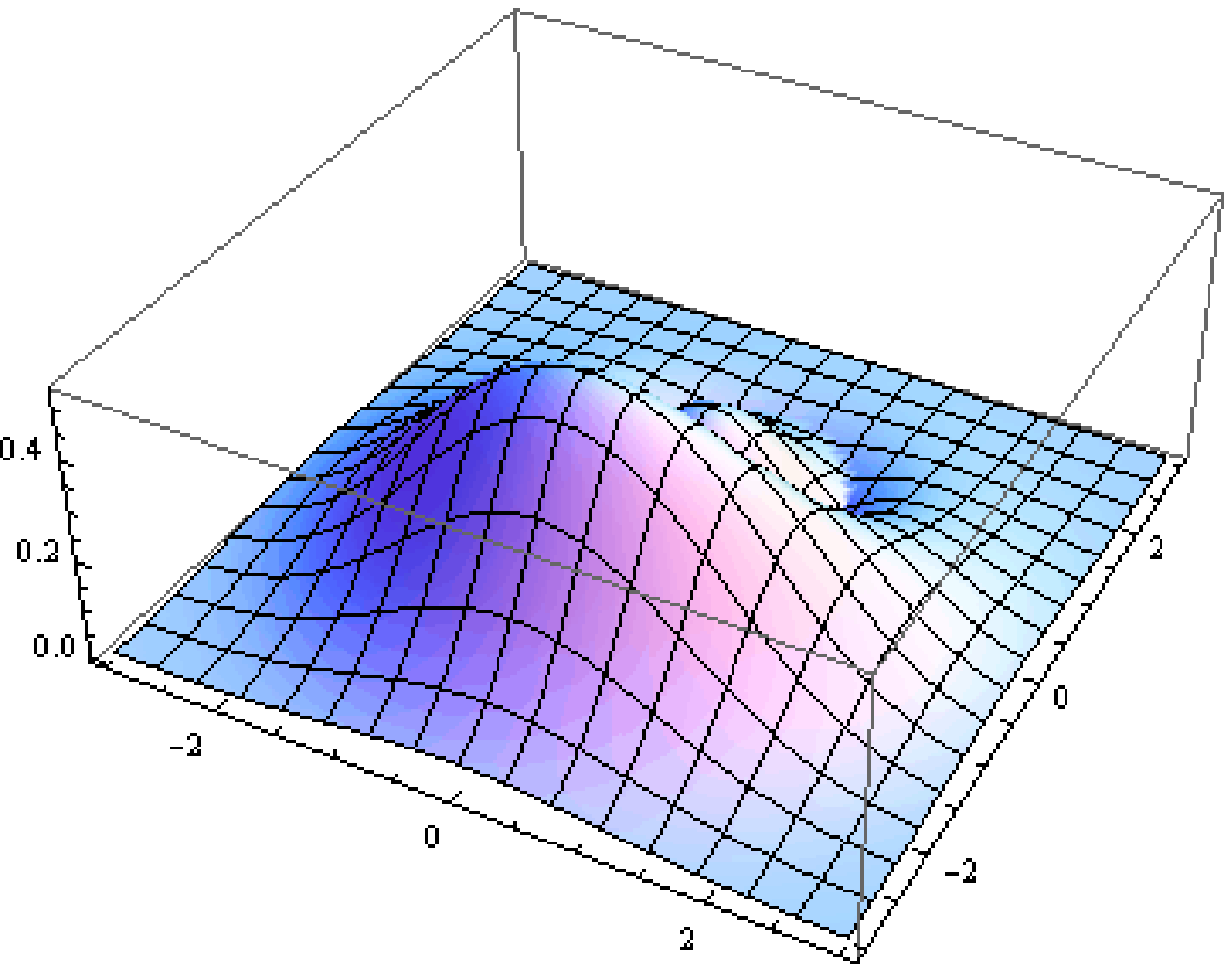,height=4cm} \\
  \epsfig{file=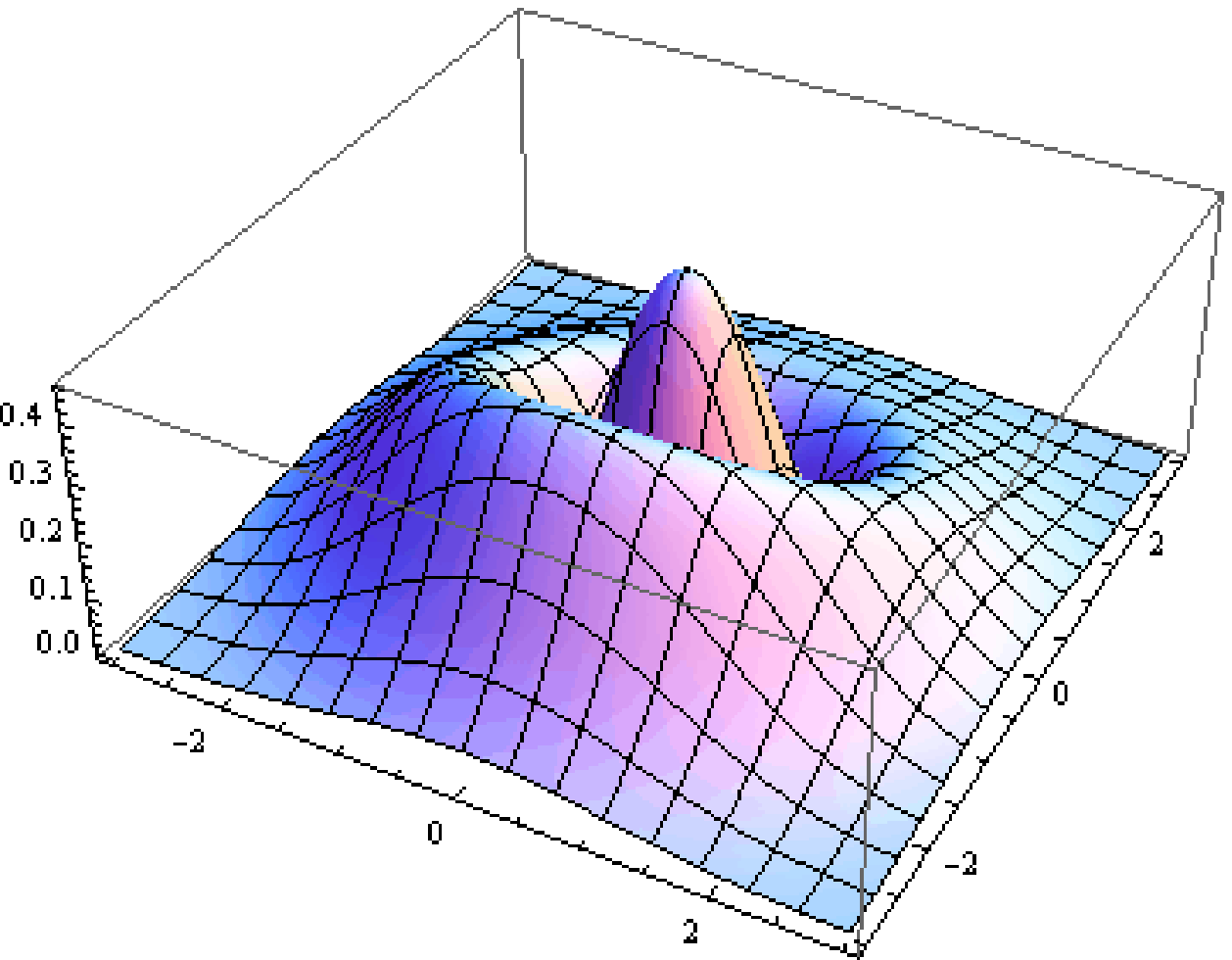,height=4cm} & \epsfig{file=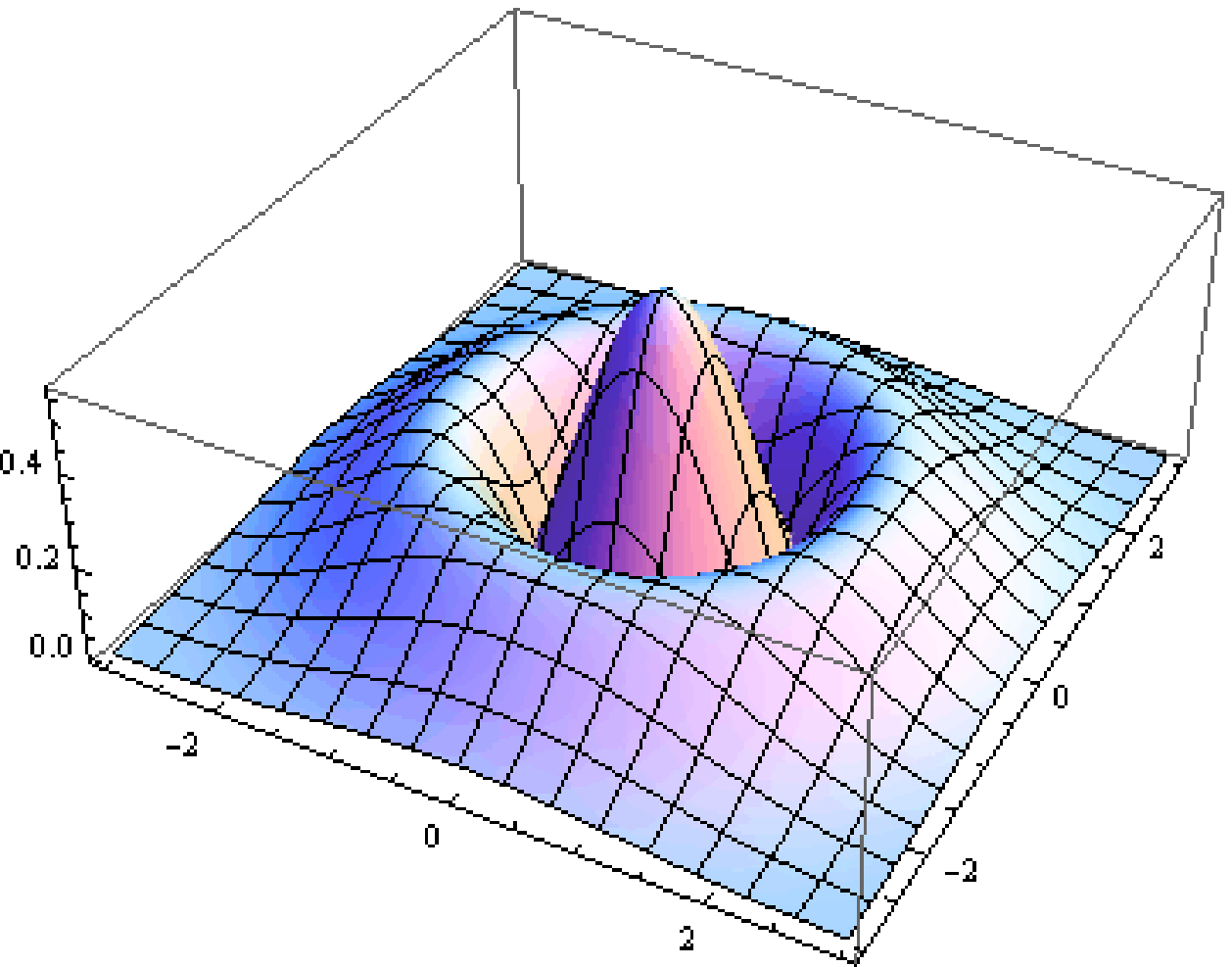,height=4cm} & \epsfig{file=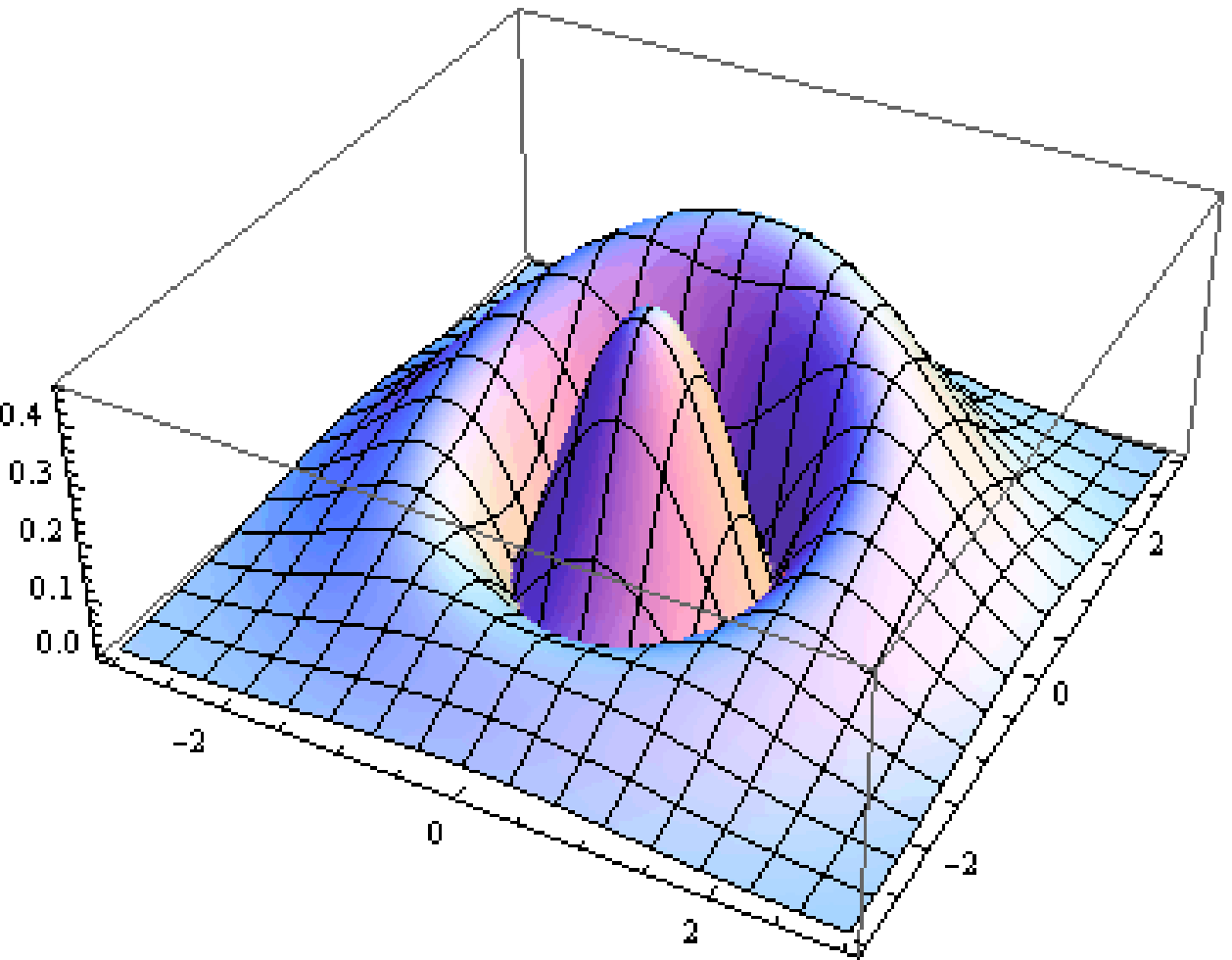,height=4cm} \\
  \epsfig{file=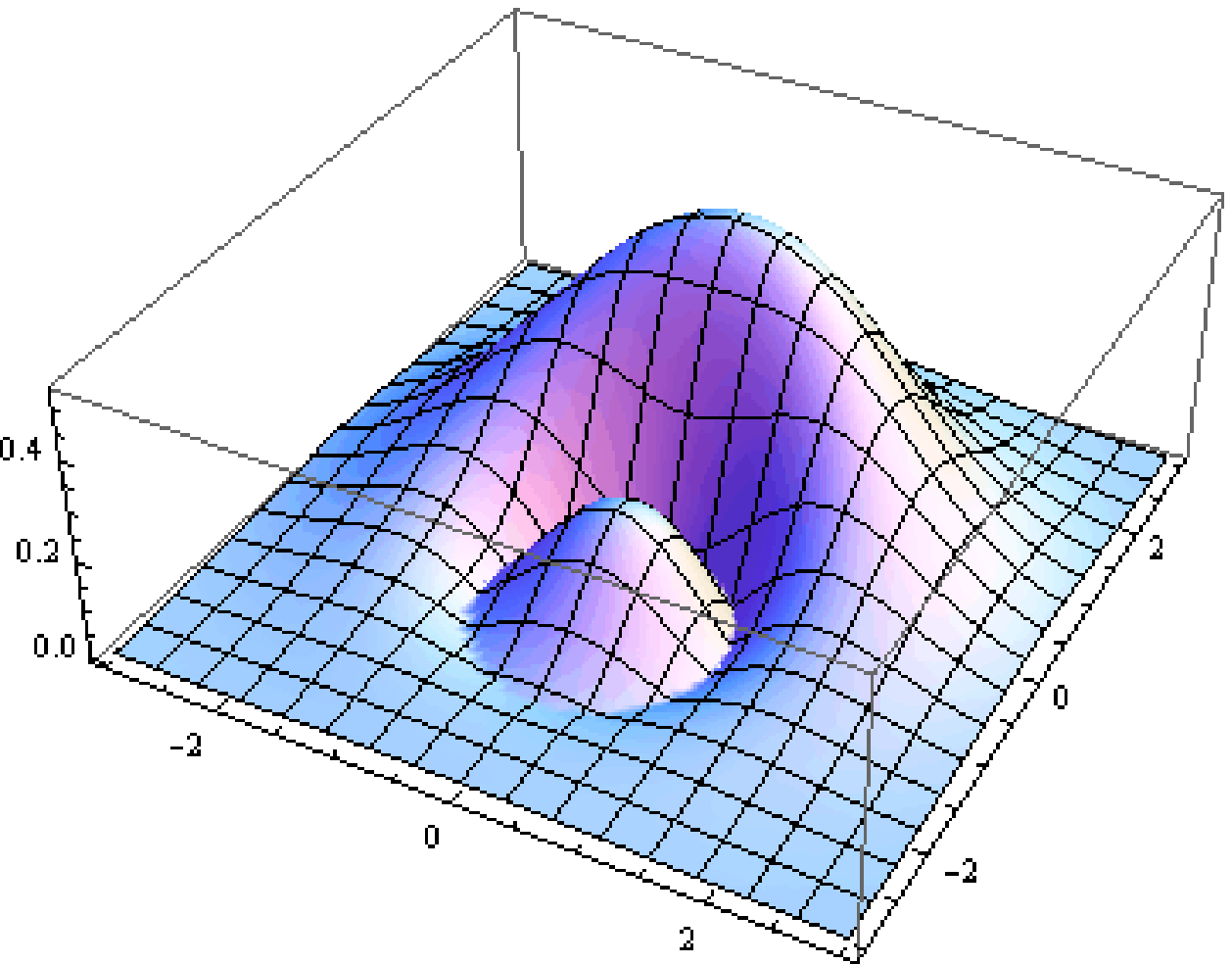,height=4cm} & \epsfig{file=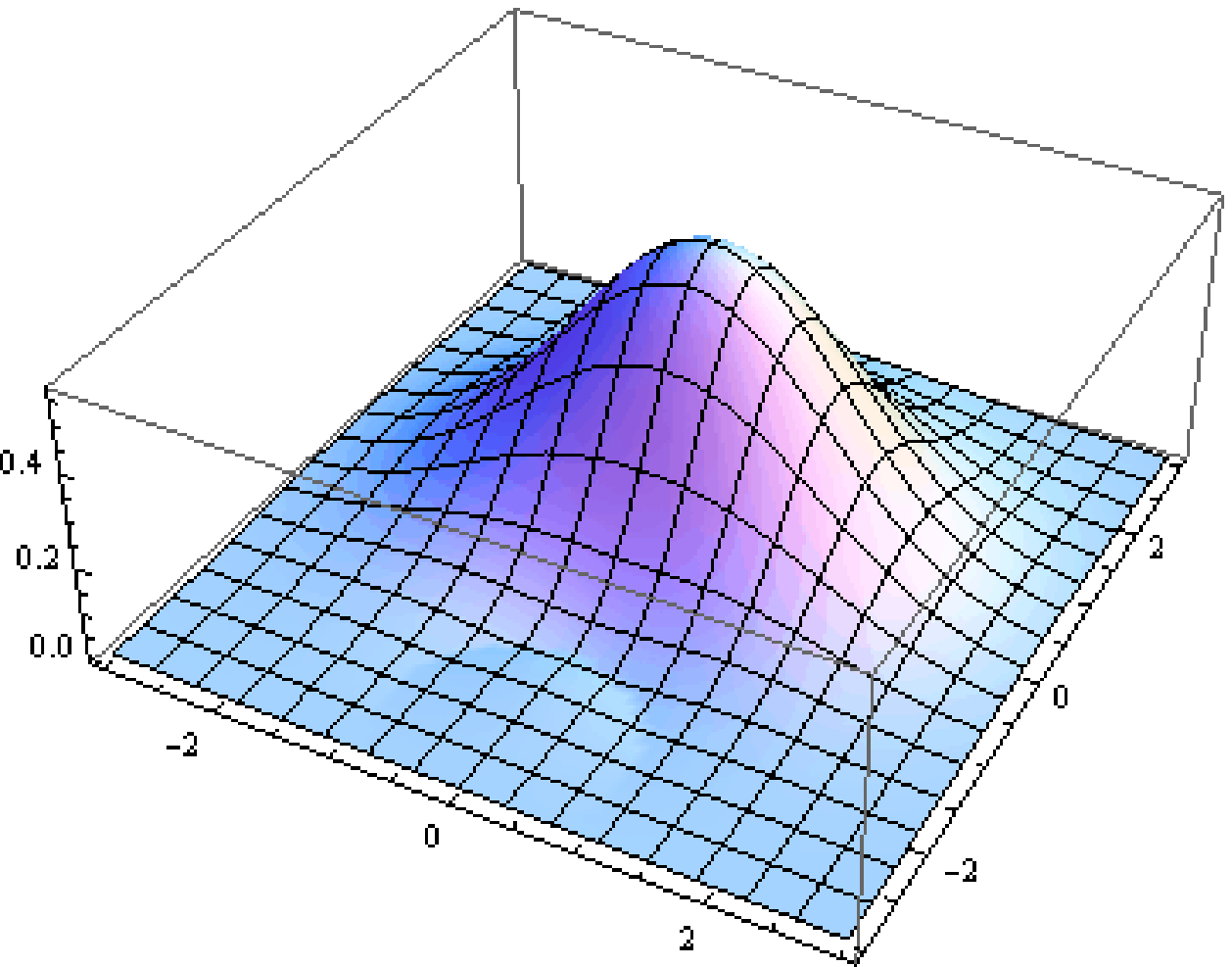,height=4cm} & \epsfig{file=L8.eps,height=4cm} \\
\end{tabular}

\caption{For convenience we let $h=1$ and $T=t/\sqrt{2}$, so that
$U_{T\sqrt{2}}^{\star}\tilde{W}h_{00}=\pi^{-1/2}e^{-(x^{2}+y^{2})/2}[\cos(2T)-y\sin(2T)+(x^{2}+y^{2})\sin^{2}(T)]$.
The figure shows the absolute value for $T=k\pi/8$, $k=0,\ldots,8$.}\label{F:FINALnineLG}
\end{center}
\end{figure}

Just as we defined the pullback of $\calt$, we may similarly define the pullback of the unitary propagator $U_{t}=e^{-it\tilt/h}$:
\begin{equation*}
    (U_{t}^{\star}\tilde{W})(f\otimes\overline{g})=\tilde{W}(U_{t}f\otimes\overline{U_{t}g}).
\end{equation*}
One may check that $U_{t}$ acts as follows:
\begin{equation*}
    U_{t}h_{0}=\cos((h/2)^{1/2}t)h_{0}+i\sin((h/2)^{1/2}t)h_{1}.
\end{equation*}
Hence we have
\begin{align*}
    U_{t}^{\star}\tilde{W}h_{00}&=\cos^{2}((h/2)^{1/2}t)\tilde{W}h_{00}+\sin^{2}((h/2)^{1/2}t)\tilde{W}h_{11}\\
    &\qquad\qquad\qquad+i\cos((h/2)^{1/2}t)\sin((h/2)^{1/2}t)[\tilde{W}h_{10}-\tilde{W}h_{01}],
\end{align*}
where, in polar coordinates,
\begin{equation*}
    \tilde{W}h_{10}-\tilde{W}h_{01}=(\pi h)^{-1/2}(r/\sqrt{h})e^{-r^{2}/(2h)}(2i\sin\theta).
\end{equation*}
This transformation is pictured in Figure \ref{F:FINALnineLG}.

As mentioned in Section \ref{S:scWig}, we have an approximate formula for the evolution of the Wigner transform in terms of the Hamilton flow. At the end of this paper, we will adapt the result to the case of the extended Wigner transform, for the study of LG mode intensities; but first we will explicitly compute the Hamilton flow.

\vspace{12pt}

The Weyl symbol of $\tilt$ is
\begin{equation*}
    p(x,\xi;h)=\frac{1}{2}x(x^{2}+\xi^{2})-2hx.
\end{equation*}
Later we will need a re-scaled version of $p$, so we consider the slightly more general symbol
\begin{equation*}
    p_{r}(x,\xi;h)=\frac{1}{2}x(x^{2}+\xi^{2})-\frac{1}{2}r^{2}hx.
\end{equation*}
Hamilton's equations for this symbol are
\begin{equation}\label{E:HamEqp0}
    \begin{cases}
    \dot{x}=x\xi\\
    \dot{\xi}=-\frac{3}{2}x^{2}-\frac{1}{2}\xi^{2}+\frac{1}{2}r^{2}h,
    \end{cases}
\end{equation}
with the conserved quantity $C=p_{r}(x,\xi;h).$

First suppose that $C\neq 0$. Then we have
\begin{equation*}
    x(t)=\frac{1}{2}C\left(\wp(t+t_{0})-\frac{1}{12}r^{2}h\right)^{-1}
    \qquad\text{and}\qquad \xi(t)=\frac{-\dot{\wp}(t+t_{0})}{\wp(t+t_{0})-\frac{1}{12}r^{2}h},
\end{equation*}
where $t_{0}$ is either an arbitrary real constant or an arbitrary real constant plus $\frac{1}{2}\omega_{1}$, the purely imaginary half-period of $\wp$. Here $\wp$ is the Weierstrass $\wp$-function associated to the invariants
\begin{equation*}
    g_{2}=\frac{1}{12}r^{4}h^{2} \qquad\text{and}\qquad g_{3}=\frac{1}{4}C^{2}-\frac{1}{216}r^{6}h^{3}.
\end{equation*}
For further details, one may consult our previous paper \cite{R:VVCP}.

\begin{figure}
\begin{center}
\epsfig{file=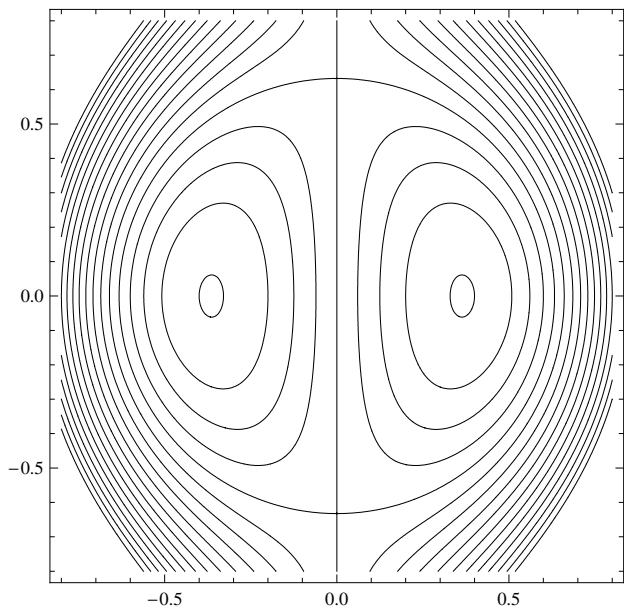,height=8cm} \caption{Hamilton flow lines in the
$(x,\xi)$ plane. Here $h=1/10$ and $r^{2}=4$.}\label{F:hflow}
\end{center}
\end{figure}

When $r=0$ and $C\neq 0$, $\xi(t)$ is always strictly decreasing, which follows simply from Hamilton's equations (\ref{E:HamEqp0}). However, when $r>0$ we have a more complicated behavior, as shown in Figure \ref{F:hflow}. There is a pocket of radius $r\sqrt{h}=\frac{r}{\sqrt{2}}w_{0}$, where $w_{0}$ is in practice the radius of the laser beam's waist.

For $C=0$, depending on the initial conditions, we have one of the following four cases:
\begin{equation*}
    \begin{cases}
    x(t)= 0\quad\forall t\\
    \xi(t)=\sqrt{r^{2}h}\coth\left(\frac{\sqrt{r^{2}h}}{2}(t+t_{0})\right),
    \end{cases}
\end{equation*}
\begin{equation*}
    \begin{cases}
    x(t)= 0\quad\forall t\\
    \xi(t)=\sqrt{r^{2}h}\tanh\left(\frac{\sqrt{r^{2}h}}{2}(t+t_{0})\right),
    \end{cases}
\end{equation*}
\begin{equation*}
   \begin{cases}
   x(t)= \pm \sqrt{r^{2}h}\sech(\sqrt{r^{2}h}(t+t_{0}))\\
   \xi(t)=-\sqrt{r^{2}h}\tanh\left(\sqrt{r^{2}h}(t+t_{0})\right),
   \end{cases}
\end{equation*}
or
\begin{equation*}
    \begin{cases}
    (x(t),\xi(t))=(0,\pm\sqrt{r^{2}h})\qquad\forall t\quad\text{(hyperbolic stationary points).}
    \end{cases}
\end{equation*}

Of course, we also have the elliptic stationary points $(x,\xi)=(\pm \sqrt{\frac{r^{2}h}{3}},0)$, corresponding to $C=\mp (r^{2}h/3)^{3/2}$.

\vspace{12pt}

Since the LG modes are extended Wigner transforms of HG modes, we now see that the intensity of LG modes evolves according to the Hamilton flow we have just computed. As discussed in Section~\ref{S:scWig}, we have that the evolution of the Wigner transform may be approximately described by the Hamilton flow $\kappa_{t}$. However, the extended Wigner transform acts slightly differently with respect to the Weyl quantization:
\begin{equation*}
    \langle\Op(\sigma)f|g\rangle=(\pi h)^{-1/2}\iint\tilde{W}(f\otimes\overline{g})(\sqrt{2}\, x,-\sqrt{2}\,\xi)\sigma(x,\xi)\,dx\,d\xi.
\end{equation*}
Even so, the same arguments give the same result, up to some minor re-scaling. Let $\Mtw$ be the operator given by
$$\Mtw(x,\xi)=(\sqrt{2}\, x,-\sqrt{2}\, \xi).$$
Then we have
\begin{align*}
    (U_{t}^{\star}\tilde{W})(f\otimes\overline{g})
    &=\tilde{W}(U_{t}f\otimes\overline{U_{t}g})\\
    &=\tilde{W}(f\otimes \overline{g})\circ\Mtw\circ\kappa_{-t}\circ\Mtw^{-1}+\mathcal{O}(h^{2})||f||_{L^{2}}||g||_{L^{2}}\\
    &=\tilde{W}(f\otimes \overline{g})\circ\tilde{\kappa}_{-t}+\mathcal{O}(h^{2})||f||_{L^{2}}||g||_{L^{2}}.
\end{align*}
Here $$\tilde{\kappa}_{t}:=\Mtw\circ\kappa_{t}\circ\Mtw^{-1},$$
which we note is the Hamilton flow of the function
\begin{equation*}
    \tilde{p}=-2p\circ\Mtw^{-1}.
\end{equation*}
Up to a re-scaling of $t$, we then have the Hamilton flow computed earlier, but now in the case $r^{2}=8$.

We are most interested in HG modes and LG modes, so we take $f\otimes\overline{g}=h_{mn}$. We then summarize our work in the following theorem:
\begin{theorem}
    For $\tilt=-2^{-1/2}h^{3/2}\calt=-2^{-1/2}h^{3/2}(\cre+\anh-\cre\cre\anh-\cre\anh\anh)$, having Weyl symbol $p(x,\xi;h)=\frac{1}{2}x(x^{2}+\xi^{2})-2hx$, and for $U_{t}=e^{-it\tilt/h}$, we have
    \begin{align*}
        (U_{t}^{\star}\tilde{W})(h_{mn})
        &\equiv \tilde{W}(U_{t}h_{m}\otimes\overline{U_{t}h_{n}})\\
        &=\tilde{W}(h_{mn})\circ\tilde{\kappa}_{-t}+\mathcal{O}(h^{2}).
    \end{align*}
    Here $\tilde{\kappa}_{t}$ is the Hamilton flow of the function $\tilde{p}$ given by
    \begin{align*}
        \tilde{p}(x,\xi;h)&=-2p(x/\sqrt{2},-\xi/\sqrt{2};h)\\
        &=-\frac{1}{\sqrt{2}}\left[\frac{1}{2}x(x^{2}+\xi^{2})-4hx\right].
    \end{align*}
\end{theorem}

\vspace{12pt}

Hence the intensities of the LG modes evolve along elliptic curves, as pictured in Figure~\ref{F:hflow}. This was in fact the motivation for the present paper. Previously, in \cite{R:VVCP}, we found that the Wigner transforms of HG modes evolve, in four-dimensional phase space $(x,y,\xi,\eta)\in\Rbb^{4}$, according to a slightly more complicated flow, where the flow in the $(x,\xi)$ plane (with $x$ and $\xi$ dual variables) for certain values of $(y,\eta)$ appears as in Figure~\ref{F:hflow}. Now we have found a situation where the flow along elliptic curves appears in the physical $(x,y)$ plane. If one wishes, one may then consider Wigner transforms of LG modes; formulas for these are given in \cite{R:VVLG}.


\vspace{12pt}

\begin{thebibliography}{20}
\small

    \bibitem{R:Alla}
        Allahverdiev, B. P.,
        ``Extensions, dilations and functional models of infinite Jacobi
        matrix,''
        Czechoslovak Math. J. 55(130), no. 3, 593--609 (2005).

    \bibitem{R:Berezanskii}
        Berezanskii, M.,
        \emph{Expansions in Eigenfunctions of Selfadjoint Operators},
        Transl. Math. Monographs 17, Amer. Math. Soc., Providence (1968).

    \bibitem{R:CalvoPicon}
        Calvo, G. F., and Pic\'{o}n, A.,
        ``Manipulation of single-photon states encoded in transverse spatial modes: possible and impossible
        tasks,''
        Phys. Rev. A 77, 012302 (2008).

    \bibitem{R:Folland}
        Folland, G. B., \emph{Harmonic Analysis in Phase Space}, Annals of Mathematics Studies,
        122, Princeton University Press, Princeton, NJ (1989).

    \bibitem{R:VVLG}
        VanValkenburgh, M.,
        ``Laguerre-Gaussian modes and the Wigner transform,''
        Journal of Modern Optics, Volume 55, Number 21, 3535--3547 (2008).

    \bibitem{R:VVCP}
        VanValkenburgh, M.,
        ``Manipulation of semiclassical photon states,''
        Journal of Mathematical Physics, Volume 50, 023501 (2009).

\end{thebibliography}
\end{document}